\documentclass[12pt]{article}
\usepackage{epsfig}
\usepackage[dvips]{color}

\begin{document}
\title{Sensitivity of the Mean Field Dynamics to the Spatial Distribution in Nuclear One-body Dissipations}

\author{\small Yong-Zhong XING$^{1}$ \footnote{E-mail:
 yzxing@tsnu.edu.cn},~Wei-Cheng FU$^{1}$, Xiao-Bin LIU$^{1}$,Fei-Ping LU$^{1}$,\\
\small  Hong-Fei ZHANG$^{1,2}$ and Yu-Ming ZHENG$^{1,3}$ }
\date{}
\maketitle

\begin{center}
$^{1}${\small Institute for the Fundamental physics, Tianshui Normal
University, Gansu, Tianshui 741000, P. R. China}\\
$^{2}${\small School of Nuclear Science and Technology, Lanzhou
University, Lanzhou 730000, P. R. China}\\
$^{3}${\small China Institute of Atomic Energy, P.O. Box 275(18),
Beijing  102413, P. R. China}\\
\end{center}

\baselineskip 0.3in.
\begin{center}{\bf Abstract}\end{center}
\hskip 0.2in
\begin{small}
\begin{quote}

Started from the static excited finite atomic nucleus, we have simulated the dynamical propagation of the nucleons using Quantum Molecular Dynamics model (QMD) without the collision term and calculated the Largest Lyapunov Exponent (LLE). The sensitivity of the nonlinear mean field dynamics to the initial conditions, in particular to the initial spatial distribution of the nucleons in certain thermodynamical stable condition, has been confirmed numerically by calculating the LLE defined in the event space. Comparatively, the Lyapunov-like exponent defined in phase space represents more clearly the sensitivity of the nonlinear dynamics to the mechanical stability of the initial conditions within the nuclear reaction time at intermediate energies. These conclusions are not only helpful for us to understand more deeply the mechanism of the nuclear reaction dynamics at intermediate energies, but it is also beneficial for clarifying further the concrete performances of deterministic chaos in the heavy-ion collisions.

\end{quote}
\end{small}

{\noindent {\bf Keywords}:  Sensitivity; Largest Lyapunov exponent; Mean field dynamics;
\\ $~~~~~~~~~~~~~~$  Exicted finite nucleus; Quantum Molecular Dynamics.}\\

{\bf PACS} number(s):24.60.Lz, 25.70.Pq

\newpage
\baselineskip 0.3in

{\section{Introduction}}

To explore the property of nuclear interaction via heavy-ion collision (HIC) at intermediate energy is a longstanding topic and has been extensively studied both theoretically and experimentally[1-8]. One of effective means to get the knowledge of the interaction is to look at the dissipation phenomena emerging in the colliding processes[9-12]. Although enormous efforts have been made and great success has been achieved, the dissipation feature as well as the dynamical mechanism of the HICs is still not completely clear so far. The difficulties for solving this problem come mainly from two aspects: one is the high complexity of nuclear interaction itself, in particular, under the extreme conditions.  The other is that the transportations and collisions of nucleons in very complicated medium environment are substantially quantum many-body dynamical processes. Since the huge degree of freedoms and the infinite dimension of the Hilbert space, from the theoretical point of view,one has to simulate the collision by means of the semiclassical prescription and evaluate some reduced variables to acquire the information of the realistic dynamical process. In such treatment both the effective interactions and the propagations of the particles involved in the process are highly nonlinear which make the dissipative phenomena more severe.

As it has well known, the dissipative phenomena relate intimately to the nonlinear dynamics or deterministic chaos which is a newly emerged science and has been developed comprehensively in recent years[13,14]. In order to learn about the particular characteristics of these dissipative processes and exploit further the mechanism of the nuclear reaction dynamics, some typical nonlinear features implicated in the course of HICs have been revealed. The power law of the size distributions of the fragments [15], the intermittency pattern of fluctuations in nuclear multifragmentation [16], and the fractal characterization [17] have been brought out quantitatively. Of course, much profound physical implications hidden behind these phenomena, such as the spinodal instability and phase transitions of the excited nuclear matter occurring in the HIC at medium energy region, have also been excavated and explained more deeply.  However, the collisions of nucleons in the HICs processes are strongly dissipative processes which cause random motion of nucleons. Correspondingly, in the theoretic simulation of the HICs, the momenta of a pair of colliding partner are assigned randomly just after their collisions [18]. Such a treatment would increase inevitably the randomness. In view of these causes, ones[19-25] have to study separately the one-body and two-body dissipations to clarify the dynamical mechanism of the HICs. Therefore, it is necessary to reveal the concrete performance of the deterministic chaos, which is a manifestation of the randomness generated by deterministic dynamics,  in the process of one-body dissipation in real reaction process.

In the present paper, based on the Quantum Molecular Dynamics model (QMD) which has frequently and successfully used to simulate the HICs at intermediate energy region, we study the one-body dissipative effects and focus on the behavior of the nonlinear dynamic characteristics of the nucleon evolution in the self-consistent mean field. Since the essential characteristic of chaos is its sensitivity to initial conditions, we study here the specific performances of the sensitivity in nuclear one-body dissipation to various conditions. To this end, in the following section we need first to prepare an initial state for a given finite nucleus under various thermodynamical conditions, and to calculate the typical nonlinear dynamical characteristics quantities in the temporal-spatial evolution of the  system under the action of the effective self-consistent mean field which can describe correctly the propagation of nucleon in real reactions.

{\section {Initialization of finite nucleus under different thermodynamical conditions}}
The potential energy of a nucleon in finite nuclei can be written in terms of the combination of the density dependent Skyme mean field
$U^{Sky}$,asymmery potential$U^{sym}$[26],the Coulomb potential$U^{Coul}$ and surface interaction $U^{Surf}$[27], i. e.,
$$
U_{q}=U^{Sky}+U^{Sym}+U^{Coul}+ U^{Surf},\eqno (1)
$$
with $q=n,p$ and
 $$
U^{Sky}=\alpha (\frac{\rho}{\rho_{0}})+ \beta (\frac{\rho}{\rho_{0}})^{\gamma},   \eqno (2)
$$
$$
U^{Sym}=\pm 2e_{a} \frac{\rho_{n}-\rho_{p}}{\rho_{n}+\rho_{p}}{\rho}     \eqno(3)
$$
$$
U^{Coul}=\frac{6}{5} Ze^{2} (\frac{4\pi\rho}{3A})^{1/3}     \eqno(4)
$$
$$
U^{Surf}=\gamma(T) (\frac{4\pi\rho}{3A})^{1/3} \rho^{-1}     \eqno(5)
$$
where $\rho=\rho_{n}+\rho_{p}$ and $\rho_{n}$, $\rho_{p}$ are  the densities of protons and neutrons in the  nucleus, respectively. $\rho_{0}=0.16 fm^{3}$  is the saturation  density of nuclear matter at temperature zero. $\gamma(T)=1.14(1+\frac{3T}{2T_{c}})(1-\frac{T}{T_{c}})$ is
 the temperature-dependent surface tension.  By virtue of the Gibbs-Duhem relation [26]
$$
\frac{\partial P}{\partial \rho}=\frac{\rho}{2}[(1+\delta)\frac{\mu_{n}}{\partial \rho} + (1-\delta)\frac{\mu_{p}}{\partial \rho} ]   \eqno (6)
$$
and the chemical potentials of nucleons
$$
\mu_{q}=U_{q}+T[\log(\frac{\lambda_{T}^{3}\rho_{q}}{2})+\sum_{}\frac{n+1}{n}b_{n}(\frac{\lambda_{T}^{3}\rho_{q}}{2})^{n} ],~~~~q=n,p   \eqno (7)
$$
it is easily to obtain the relationships of the pressure, density  and temperature of the system. $\lambda_{T}=(\frac{2\pi\hbar^{2}}{mT})^{1/2}$
 is the thermal wavelength of the nucleon at temperature $T$  and the sum is taken up to $n=5$ in Eq.(7) in the following calculations. Ensuring the correctness of our calculations, the finite nucleus $^{208}Pb$  is chosen in the present paper due to its thermodynamical properties had been extensively studied in some literatures and the extension of the method used here to other nuclei is straightforward. The variation of the pressure with respect to the density at different temperature $T$ for this nucleus determined with the expressions mentioned above is depicted in Fig.1. The mechanical instable, i. e, spinodal region, is indicated below the dotted curve region in this figure.

\begin{center}
{\bf Fig.1}
\end{center}

Referring to the Fig.1, the distribution of the nucleons in the initial excited nucleus, at a set of certain values of density, pressure and temperatures,  can be  obtained by the following procedure: the position of each nucleon can be sampled uniformly in a fermi-sphere, while  the momentum of it can be sampled according to the Fermi-Dirac distribution at a certain chemical potential.
$$
f(p)=\frac{1}{ 1+exp \{ \beta ( \epsilon_{q}-\mu_{q} ) \} }    \eqno  (8)
$$
Where $\beta=\frac{1}{kT}$  is Boltzmann temperature. The single particle energy is
$$
\epsilon_{q}=\frac{2\hbar^{2}}{20m} (\frac{ 3\pi^{2}\rho}{2})^{2/3}
 [ ( 1+\delta )^{5/3}+( 1-\delta )^{5/3} ]
 + \frac{\alpha}{2} ( \frac{\rho}{\rho_{0}})
+\frac{\beta}{1+\gamma}( \frac{\rho}{\rho_{0}})^{\gamma}
+\frac{C}{2}( \frac{\rho}{\rho_{0}})\delta^{2}                                \eqno  (9)
$$
The result is shown in Fig.2 with $\rho(t=0)=0.11fm^{-3}$ and $T=6MeV$.

\begin{center}
{\bf Fig.2}
\end{center}

{\section {Spatio-temporal evolution of the system}}
Based on the above preparation, we proceed to investigate the consequence of the dynamical evolution of the system. As is well known,  the Quantum Molecular Dynamics model (QMD) assumes a Lagrangian with a non-relativistic kinetic energy and a potential energy from effective interactions of nucleons. The equation of the motion of the nucleons is derived from a variational principle with the Lagrangian. The single-particle wavefunction is assumed to be a Gaussian wave-packet with a fixed width. Therefore, the correlations have been kept in the QMD and thus this model is suitable not only for studying the fragmentation of HICs at medium energies, but also for exploring the properties of the dissipation of the dynamical process. The main ingredients of the QMD have been given repeatedly in a great deal of literatures, see for example refs. [28-32] and references therein. Here we recall briefly the mean field involved in this model only.
$$
U_{q}=U^{Sky}+U^{Sym}+U^{Coul}+ U^{Yuk}+ U^{Pauli},  \eqno (10)
$$
where the expresses of $U^{Sky}$ and $U^{Sym}$ are the same as given in the eq.(2).  The other terms in eq. (10) are the potentials folded with the Gaussian wave packets which represent the nucleons in QMD model and read respectively as:
$$
U^{Coul}=\frac{e^{2}}{4}\sum_{ j \ne i} \frac{(1-t_{z_{i}})(1-t_{z_{i}})[1-erfc(r_{ij}/\sqrt{4L})]}{r_{ij}}                    \eqno(11)
$$
$$
U^{Yuk}=\frac{t_{3}}{2}\sum_{ j \ne i} \frac{1}{r_{ij}}exp(-a r_{ij})erfc(a\sqrt{L}-r_{ij}/\sqrt{4L})              \eqno(12)
$$
$$
U^{Pauli}=V_{p}(\frac{\hbar}{p_{0}q_{0}})^{3}
exp\{ -\frac{(r_{i}-r_{j})^{2}}{2q_{0}^{2}}-\frac{(p_{i}-p_{j})^{2}}{2p_{0}^{2}}      \}
\delta_{p_{i}p_{j}}                                 \eqno(13)
$$
All the parameters involved in these equations are list in the Refs. [33,34].

\begin{center}
{\bf Fig.3}
\end{center}

Although the temporal variation of the density fluctuation of hot nuclear matter produced in nuclear reaction processes has been studied by many authors,e.g., Ref.[35,36], we calculate here the variation of the density and its fluctuation for the system at certain thermodynamical conditions since our purpose being to observe the non-linear dynamical features of the evolution of the deterministic system, or the chaotic characters, which is considered as a long time behavior of the non-linear dynamics. Correspondingly, the relative variations of the averaged density $<\rho(t)>/\rho(t=0)=\frac{1}{\rho(t=0)} \int \rho^{2}(t) d\tau$ and the fluctuation $\sigma= \frac{<\rho^{2}(t)>-<\rho(t)>^{2}}{<\rho(t)>^{2}}$ of the system starting from different initial conditions are plotted in the up panel of Fig. 3. The solid line corresponds to the  $\rho(t=0)=0.11fm^{-3}$ and the dotted line stands for $\rho(t=0)=0.06fm^{-3}$ at temperature $T=6MeV$.  The former is in the stable region of the nucleus and the latter is unstable mechanically as shown in Fig. 1. The up panel of Fig.3 shows that the overall behavior of the relative density fluctuation increases continually in the course of the system evolution for both initial states. This means that the increase of the density fluctuation is a common performance for the excited finite nucleus not only located in the mechanical instable zone, but also for the system adjacent to the zone. However, the discrepancies between them are still obvious. Namely, the amplitude of the increasing of the former is larger than that of the latter. This is just the different consequences of the initial states under the various conditions of mechanically stability and instability. The averaged density profile at time $t=800 fm/c$  is plotted in the below panel of the Fig.3. Where the horizontal axis of the figure represents the coordinates of the nucleons in radial direction. Although there isn't inhomogeneity of the distribution in the radial direction can  been displayed apparently in the diagram due to spatial averages have been taken in the calculations, the differences are obvious by comparing the curves started from different initial states. This distinction can be attributed to the difference of their initial densities in nuclear physics.  In view of nonlinear dynamics, however, this means the sensitivity of the system evolution to its initial condition which is the essential characteristic of deterministic chaos. Therefore, we proceed to explore the nonlinear dynamics manifestation of the spatial-temporal evolution of the highly nonlinear dynamical system in the next section.

{\section{The Largest Lyapunov exponent}}

 The Largest Lyapunov Exponent (LLE)[18,37] is a characteristic quantity which can be used to quantify chaotic motion of nonlinear  dynamics.  Basically, it is defined as
$$
\lambda=\lim_{t\to \infty} \lim_{d(0) \to 0}\frac{1}{t} \log\frac{\| d(t) \|}{\| d(0) \|}              \eqno(14)
$$
with $||d(t)||
$ being the distance of two adjacent orbits or trajectories ¡®1¡¯ and ¡®2¡¯  at a given time in phase space and $\| d(0) \|$ its initial value.  A positive LLE implies the occurrence of chaos. Specifically,
 if the $\lambda$ is greater than zero in the evolution of the system, the dynamical system is deterministic chaotic. otherwise the system is integrable and the temporal and spacial evolution of the system is regular.

For the nuclear reaction, as has stated above, the complexity existed in the issue comes not only form the high nonlinearity of the interaction of nuclear matter, but also from the description for the many-boy dynamical process. Therefore, there are various definitions for the distance or metric, $\|d(t)\|$. For instant, in Ref.[38] $\|d(t)\|$ is taken as the difference of two trajectories in norm between their density profiles while in the Refs.[39-41] it is defined as the geometric distance of  the real trajectories of part nucleons in coordinate space. Recently, Refs.[35,36] have proposed the following form:

$$
  \|d(t) \| = \sqrt{\sum_{i=1}^{N} [ ((\vec{r}_{i1}-\vec{r}_{i2})/r_{rms})^{2}-((\vec{p}_{i1}-\vec{p}_{i2})/p_{avp})^{2}] }              \eqno(15)
$$
Where the vectors $(\vec{r},\vec{p})$ is the coordinate and momentum of nucleon $i$ in phase space. The $r_{rms}$  and $p_{avp}$ are the rms radius and the average momentum of the nucleus containing nucleon $i$. The sum in eq. (15) runs over all the nucleons involved in the system. The subscripts 1 and 2 refer to two different events with the initial distance $\|d(0)\| \leq 10^{-7}$, namely the indexes $i$ here refers to the coordinates of a same nucleons in two different running. Such definition is relatively more suitable for the study of finite nuclei comparing to infinite nuclear matter due to the limitation of the particles numbers. In view of the existence of the chaos in nuclear reactions has been found by the literatures cited above and the issue with debate nowadays is focusing on the time scale on the emerging of the chaos since there are no chaos at the beginning stage of the reaction process according to the linear response theory[17], so we plot the $\lambda(t)=\frac{1}{t} \log\frac{\| d(t) \|}{\| d(0) \|}$  as a function of time $t$ in the propagation of the system with QMD in Fig.4.

\begin{center}
{\bf Fig.4}
\end{center}

In Fig.4 curves from (b) to (g) correspond respectively to the different initial densities, i.e, from  $\rho(0)=0.14 fm^{-3}$(b) to $\rho(0)=0.02 fm^{-3}$ (g) at $T=6 MeV$. The result for the saturated density $\rho(0)=0.16 fm^{-3}$ denoted by symbol (a) at zero temperature is also plotted. For each initial density the distance of the two events are taken as  $\|d(0)\| \leq 10^{-7}$.  As a comparison, the horizontal line at $\lambda=0$  is also drawn in this figure. We can see form this figure that the corresponding $\lambda(t)$ is greater than zero when the system deviates from saturation state. It is also shown that the $\lambda(t)$ corresponding to the trajectory starting from the mechanical unstable region is significantly greater than that from the mechanical stable region. When the distribution approaches to the saturated nuclear density, $\rho(0)=0.16 fm^{-3}$, the corresponding $\lambda(t)$ is less than and close to zero. This can be expected not only by the relative stability of the ground nucleus in the nuclear physics theory, but also by the insensitivity of the mean field dynamics to the initial value in the point of nonlinear dynamics view, i. e., the regular motion of the dynamics. We can also see that the discrepancies between $\lambda(t)$ values are apparent in the range of time $t\leq 100fm/c$  and after that period it will decrease with the time increasing, which reflects the weakening of interplay between different parts of the system.  These characteristics fully confirm that the LLE defined in events space is a good physical quantity for describing the spatio-temporal evolution of finite hot nucleus. Its variation with time not only shows the nonlinear chaos characteristics of one-body dissipation in the process of heavy ion collision at medium energy, but also further confirms the chaotic mechanism of multifragmentation.

Inspired by the connotation of the Lyapunov exponent, we can defined an quantity in the phase space for a given nucleus to reflect the expansion of excited finite nucleus. Specifically, we define the following Lyapunov-like exponent
$$s(t)=\frac{1}{t}\log\frac{\| D(t) \|}{\| D(0) \|}   \eqno(16)$$
to describe the expansion of the system. Where
$$ \|D(t)\| = \sqrt{ \sum_{i,j=1}^{N} [ ((\vec{r}_{i}-\vec{r}_{j})/r_{rms})^{2}-((\vec{p}_{i}-\vec{p}_{j})/p_{avp})^{2}]}      \eqno (17) $$
and the subscripts  $i$ and $j$ denote the different nucleons in a certain nucleus instead of the different events. Then we can directly look at the variation of this quantity in the process of hot system expansion to explore the sensitivity of the nonlinear dynamical system on its initial conditions. The results for two different initial densities, $\rho(0)=0.04fm^{-3}$ and $\rho(0)=0.14fm^{-3}$, are showing in Fig. 5.

\begin{center}
{\bf Fig.5}
\end{center}

Keeping the relation $D(t)=D(0)exp^{\{s(t)t\} }$ in mind, we can clearly see in this figure that the disassociation of the system in phase space is exponentially when it situates in spinodal region and the speed of its separation
dependent sensitively on the mechanically stability of the initial conditions at given temperature. That is, for the initial distribution  $\rho(0)=0.04 fm^{-3}$  at $T=6 MeV$, corresponding to the mechanical unstable region of the system initially, the separation rate, $s(t)$, is positive at least within the time scale of nuclear reactions at intermediate energies($t \leq 300fm/c$), while for $\rho(0)=0.14fm^{-3}$, the $s(t)$ at the same temperature is negative in the same time range. In this sense we can consider the definitions $\lambda (t)$ and $s(t)$ being equivalent. It is also interesting to notice, comparing the Fig.5 to Fig.4, that the $s(t)$ is negative in the first stage of the evolution of the system  when the system located in unstable region mechanically while the $\lambda (t)$ is positive as long as the density being below the saturated value ($\rho_{0}=0.16 fm^{-3})$. This difference maybe suggest us that the $s(t)$ is more effective for us to describe the expansion of hot finite nucleus. Moreover, as has mentioned above that the LLE, $\lambda (t)$, proposed in the Refs.[35,36] and its behaviour has been studied by the authors as a function of temperature in hot nuclear matter. Here our calculations reveal further the properties of the LLE varying with the initial state under different thermodynamical stable conditions at certain temperature in a excited finite nuclei. From the appearance of the Lyapunov-like exponent showing in the Fig.5, however, we can speculate that the $s(t)$ is more suitable for studying the properties of Mixing and ergodicity of the nonlinear dynamical systems. The considerations in this direction are ongoing now.

{\section{Discussion and Summary }}

Up to now, profound understanding for the dynamical mechanism of nuclear multifragmentation in the heavy ion collisions at intermediate energies has been made and some important relations between the nonlinear dynamical characteristics and the fragmentations have been proposed quantitatively [12,42,43]. However, many problems need still to be further studied due to the complexity of the issue itself. For example, it seems to be that we have known the existence of chaotic mechanism in the reaction dynamics so far, but what is the concrete manifestation of the nonlinear dynamical characteristics in the realistic reaction process rather than in a simplified one-dimensional or two-dimensional model?  How does the nonlinear feature affect the final state of the reaction?  What is the relationship between the nuclear reaction process taking place in a limited time and the deterministic chaos which is considered as a dynamical consequence of nonlinear system in long time evolution?  The answer to these issues is making sense not only to the nuclear physics or astrophysics but also very important to the nonlinear dynamics or deterministic chaos.
The motivation of this paper is just to study the characteristics of nonlinear dynamics in the one-body dissipation of excited finite nucleus. Started from the certain distribution with different mechanical instability condition, we have simulated the dynamical propagation of the nucleons in the nuclear system by using the quantum molecular dynamics model with the Skyrme-type mean field without collision term. The model and the interaction are repeatedly used to simulate the realistic nuclear reaction and reproduce the multifragmentation with great successes. Then the largest Lyapunov exponent has been calculated in the evolution of the system. The sensitivity of the nonlinear mean dynamics to the initial conditions, in particular to the initial spatial distribution of the nucleons at certain thermodynamical conditions, has been confirmed numerically. Inspired
by the definition of Lyapunov exponent, a Lyapunov-like quantity has been defined and examined. The results show that the latter can describe quantitatively the expansion of hot limited nuclear systems. Comparatively, the Lyapunov-like exponent defined in phase space can more clearly reflect sensitivity of the nonlinear dynamics system to the mechanical stability of the initial conditions during the nuclear reaction periods. These results are not only helpful for us to understand more deeply the mechanism of the nuclear reaction dynamics, but also beneficial for us to clarify further the concrete performance of deterministic chaos in the heavy-ion collisions.

\vskip 1.0cm

\section*{Acknowledgements}

We would like to thank Prof. F. S. Zhang and Dr. Wen-Fei Li
for their help and inspiring discussions. This work is supported by  the National Nature Science Foundation of
China under Grant Nos.11665019, 11764035, 11265013.

\newpage

\begin{figure}[htpb]
\centerline{ \epsfig{file=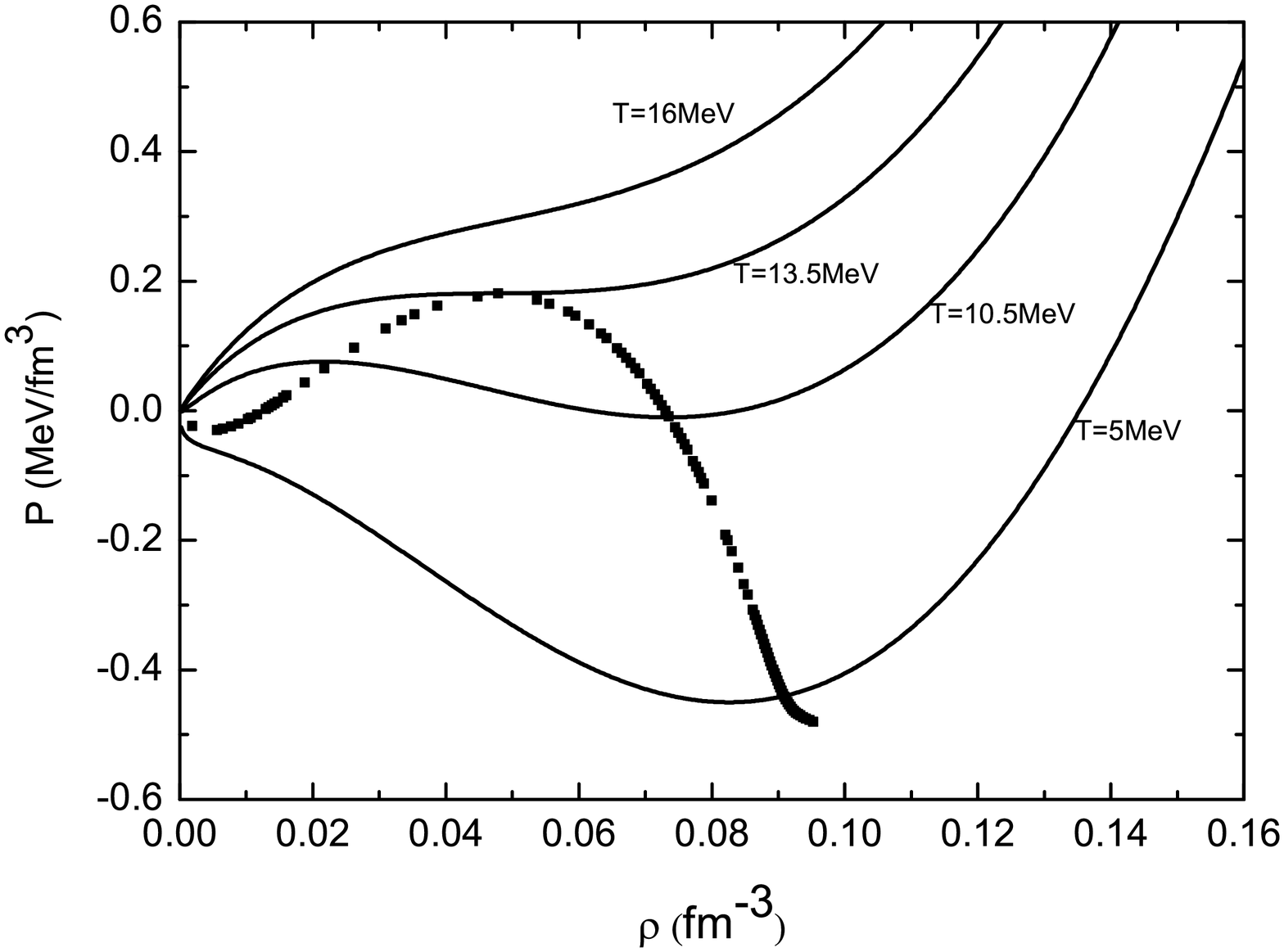,width=12cm,height=10cm,angle=0} }
\caption{The Phase diagram for the nucleus $^{208}Pb$ }
 \label{Fig1}
\end{figure}

\begin{figure}[htpb]
\centerline{ \epsfig{file=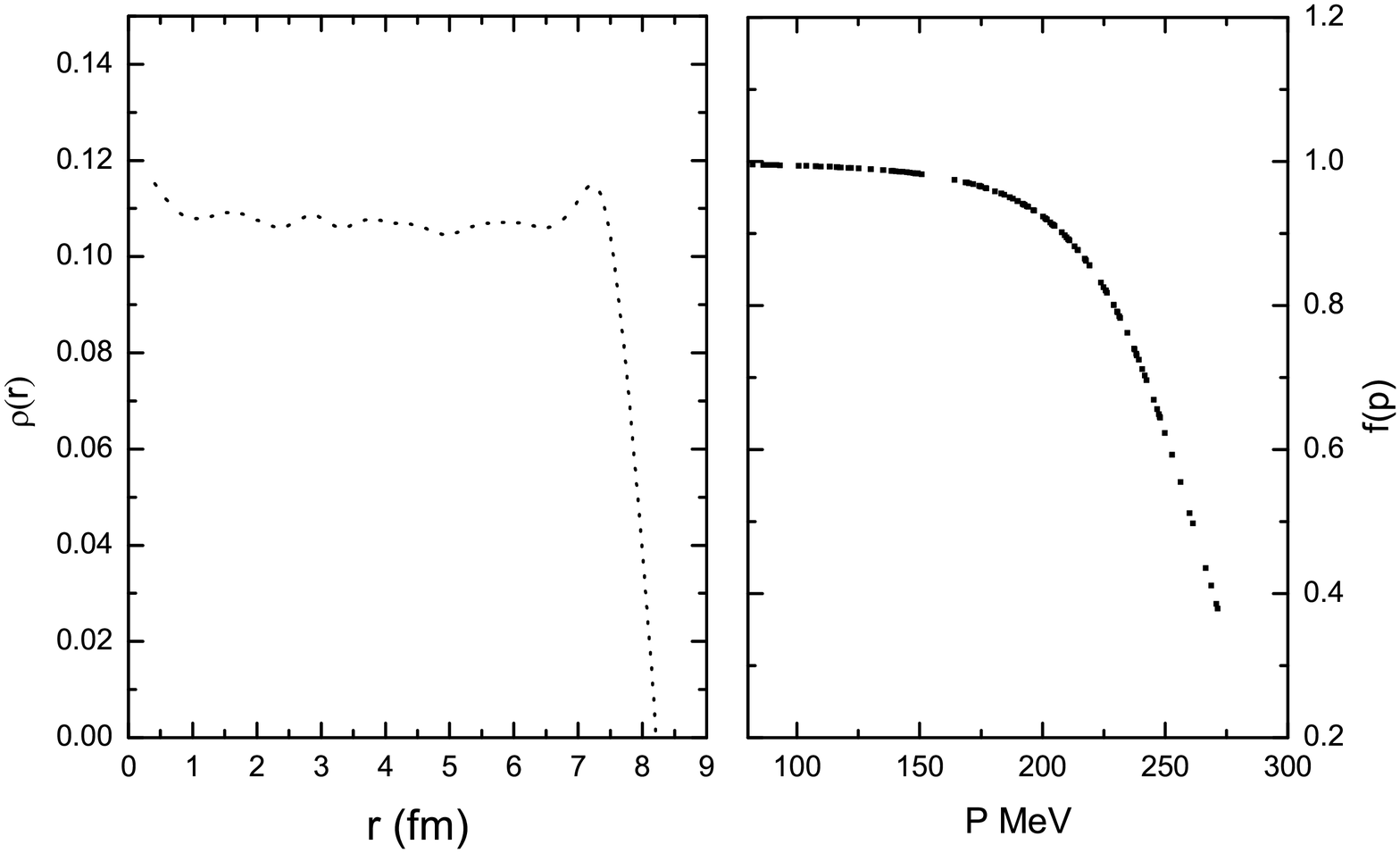,width=12cm,height=10cm,angle=0} }
\caption{Coordinates and Momenta of nucleons sampled for the nucleus $^{208}Pb$
 at $\rho(0)=0.11 fm^{-3}$ and $T=6MeV$ }
\label{Fig2}
\end{figure}

\begin{figure}[htpb]
\centerline{ \epsfig{file=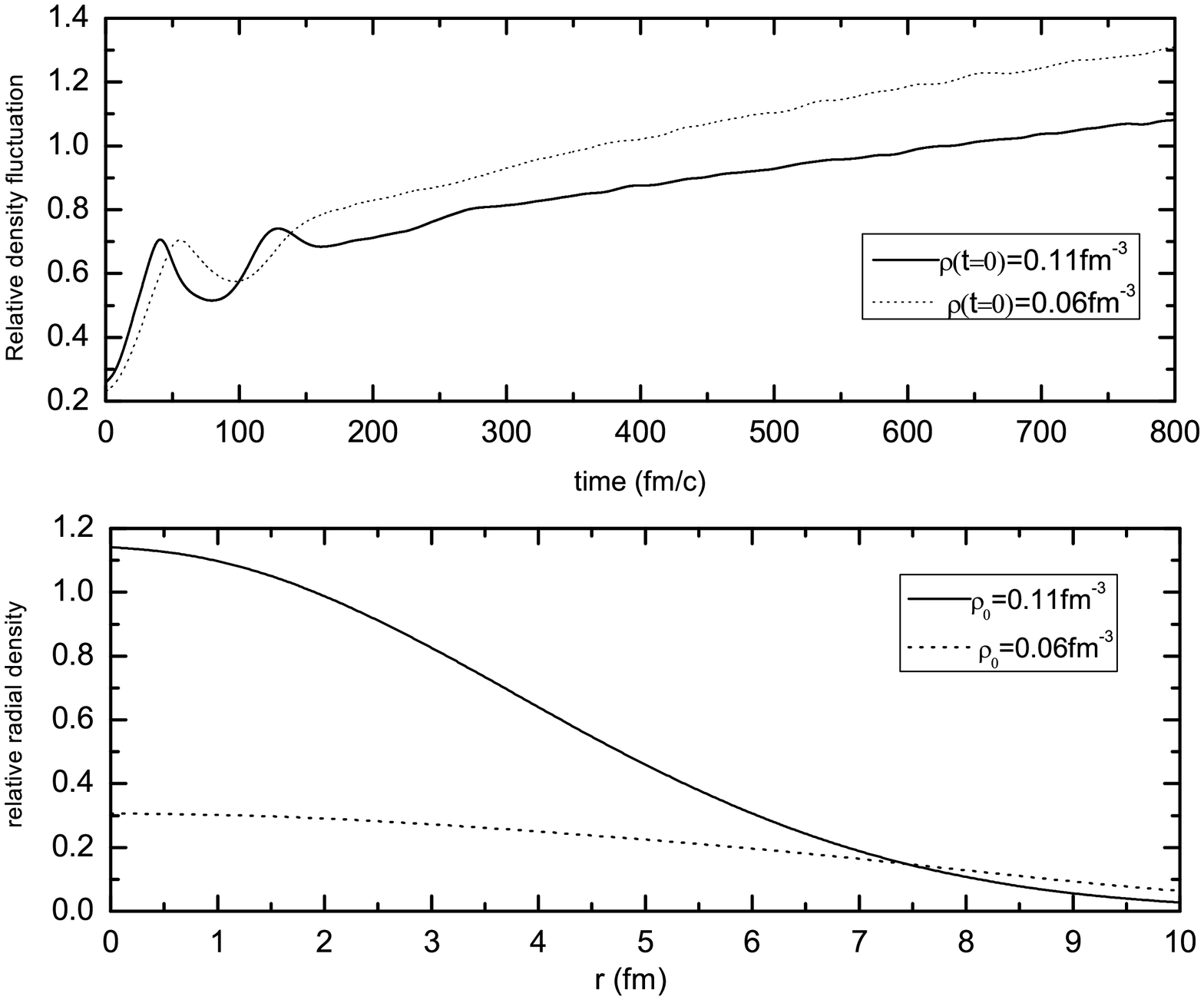,width=12cm,height=10cm,angle=0} }
\caption{The relative density fluctuation vs time (up panel)
 and the radial density profile at time $800fm/c$ (down panel) }
\label{Fig3}
\end{figure}

\begin{figure}[htpb]
\centerline{ \epsfig{file=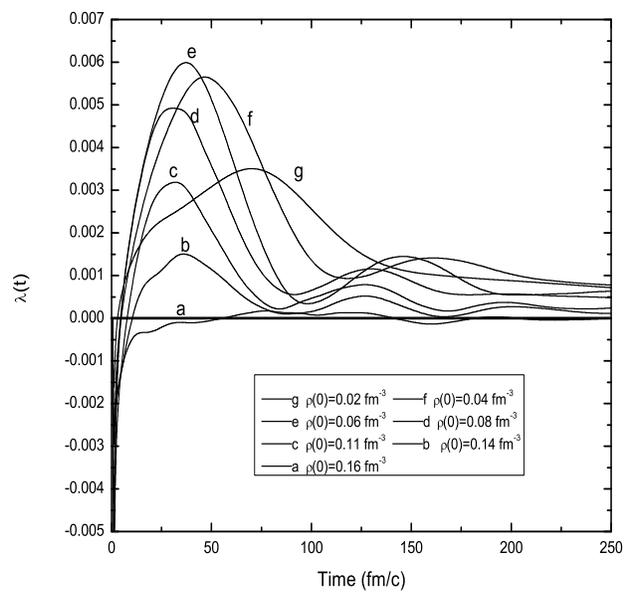,width=12cm,height=10cm,angle=0} }
\caption{ Variation of Largest Lyapunov Exponent with time for different initial states at fix temperature $T=6 MeV$.}
\label{Fig4}
\end{figure}

\begin{figure}[htpb]
\centerline{ \epsfig{file=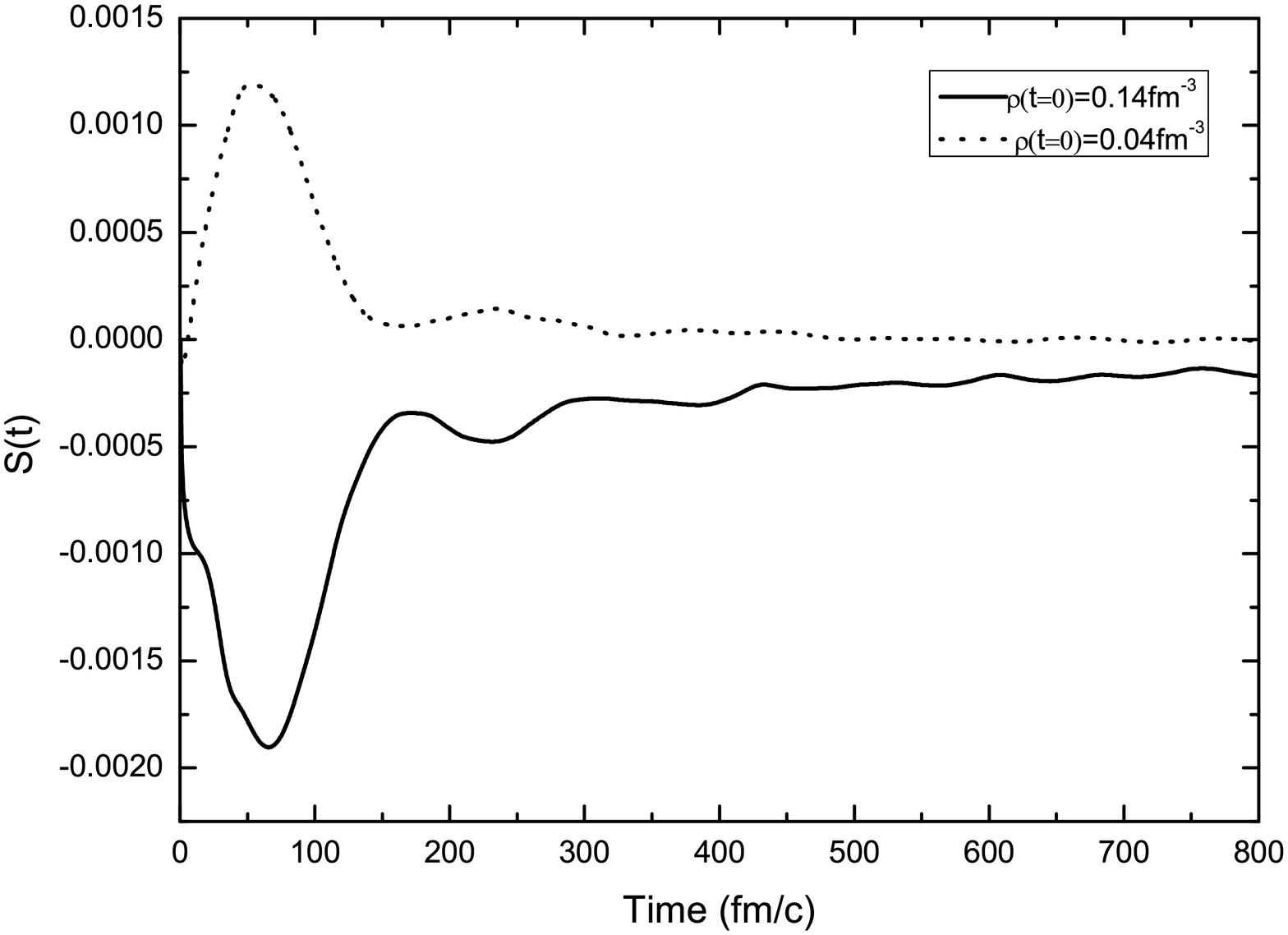,width=12cm,height=10cm,angle=0} }
\caption{ Temporal evolution of the distance between different nucleons in phase space for the system}
\label{Fig5}
\end{figure}

\end {document}